\documentclass{aa}
\usepackage{graphicx}
\usepackage{times}
\begin{document}

\title{Multiperiodicity in the large-amplitude rapidly-rotating $\beta\,$Cephei
star HD\,203664\thanks{Based on data gathered with the 1.2m Mercator telescope
equipped with the photometer P7 situated at Roque de los Muchachos, La Palma}}

\author{C.\ Aerts\inst{1,2}, P.\ De Cat\inst{3}, J. De
  Ridder\inst{1,}\thanks{Postdoctoral Fellow of the Fund for Scientific Research
  of Flanders (FWO)}, H.\ Van Winckel\inst{1}, G.\ Raskin\inst{1,4}, G.\
  Davignon\inst{1,4}, K.\ Uytterhoeven\inst{1,4,}\thanks{Present address: Center
  for Astrophysics, University of Central Lancashire, Preston PR1 2HE, UK}}

\institute{Instituut voor Sterrenkunde, Katholieke Universiteit Leuven,
Celestijnenlaan 200 B, B-3001 Leuven, Belgium \and
Department of Astrophysics, Radboud University Nijmegen, PO Box 9010, 6500 GL
Nijmegen, the Netherlands\and
Koninklijke Sterrenwacht, Ringlaan 3, 1180 Brussel, Belgium\and
Mercator Telescope, Calle Alvarez de Abreu 70, 38700 Santa Cruz de La Palma,
Spain 
}

\authorrunning{Aerts et al.} \titlerunning{Multiperiodicity in the
  large-amplitude $\beta\,$Cephei star HD\,203664} \date{Received ; accepted}

\abstract{}{We perform a seismic study of the young massive $\beta\,$Cephei star
HD\,203664 with the goal to constrain its interior structure.}  {Our study is
based on a time series of 328 new Geneva 7-colour photometric data of the star
spread over 496.8 days.} {The data confirm the frequency of the dominant
mode of the star which we refine to $f_1=6.02885\,$c\,d$^{-1}$.  The mode has a
large amplitude of 37\,mmag in V and is unambiguously identified as a dipole
mode ($\ell=2$) from its amplitude ratios and non-adiabatic computations.
Besides $f_1$, we discover two additional new frequencies in the star with
amplitudes above $4\sigma$: $f_2=6.82902\,$c\,d$^{-1}$ and
$f_3=4.81543\,$c\,d$^{-1}$ or one of their daily aliases.  The amplitudes of
these two modes are only between 3 and 4\,mmag which explains why they were not
detected before. Their amplitude ratios are too uncertain for mode
identification. } {We show that the observed oscillation spectrum of HD\,203664
is compatible with standard stellar models but that we have insufficient
information for asteroseismic inferences.  Among the large-amplitude
$\beta\,$Cephei stars, HD\,203664 stands out as the only one rotating at a
significant fraction of its critical rotation velocity ($\sim 40\%$).
} \maketitle

%\addtocounter{table}{1}

\section{Introduction}

The $\beta\,$Cephei stars are a homogeneous group of oscillating B0--B3 stars
that have been studied as a class for more than a century now. Stankov \&
Handler (2005) recently compiled an overview of the observational properties of
this group of stars.  The oscillations of $\beta\,$Cephei stars are explained in
terms of the $\kappa$\,mechanism operating in the ionisation layer of the
iron-peak elements 
{
(Cox et al.\ 1992; Kiriakidis et al.\ 1992; 
Moskalik \& Dziembowski 1992; Dziembowksi \& Pamyathnykh 1993).}
Given that mainly low-degree low-order pressure and gravity modes are excited,
these stars are good potential targets for in-depth seismic studies of the
interior structure of massive (i.e.\ pre-supernova) stars.  Indeed, the
luminosity classes of the known $\beta\,$Cephei stars range from V up to I
(Stankov \& Handler 2005; see also Waelkens et al.\ 1998) and theory predicts
the occurrence of oscillations for this whole area in the HR diagram (Pamyatnykh
1999).

Recent progress in the seismic interpretation of selected $\beta\,$Cephei stars
was remarkable in the sense that standard stellar structure models are unable to
explain the oscillation data for the best-studied stars: HD\,129929 (Aerts et
al.\ 2003), $\nu\,$Eridani (Pamyatnykh et al.\ 2004, Ausseloos et al.\ 2004),
12\,Lacertae (Handler et al.\ 2005, Ausseloos 2005). Pamyatnykh et al.\ (2004)
have suggested to include radiative diffusion processes in a new generation of
stellar models in an attempt to resolve the excitation problem in
$\nu\,$Eridani. This has not yet been achieved so far. These three well-studied
$\beta\,$Cephei stars are all slow rotators with vsini below 40\,km\,s$^{-1}$
and a mass between 9 and 12\,M$_\odot$.

In view of these recent achievements, and in an attempt to obtain similar
results for a star with higher surface rotation velocity, we have selected one
of the most rapid rotators among the large-amplitude $\beta\,$Cephei stars for a
long-term photometric monitoring programme on which we report here.

The star HD\,203664 (B0.5V, m$_{\rm V}$=8.59) was discovered to be a new
$\beta\,$Cephei star by Aerts (2000), who derived one frequency of
6.0289\,c\,d$^{-1}$ from the HIPPARCOS photometry. This frequency was confirmed
by her in 49 Geneva measurements spread over about one year taken with the P7
photometer attached to the 0.7m Swiss telescope at La Silla observatory. The
scarce multicolour data set did not allow discrimination between $\ell=1$ or 2
for the spherical degree of this oscillation mode but did seem to exclude a
radial mode. HD\,203664 is among the top ten of the class as far as photometric
amplitude is concerned, with a value of $\sim$\,30mmag in the Geneva V filter
(Aerts 2000). It is by far the most rapid rotator among the large-amplitude
members (see Fig.\,4 of Stankov \& Handler 2005), with vsini=200\,km\,s$^{-1}$
derived from high-resolution spectra by Little et al.\ (1994).  HD\,203664 also
happens to be one of the very few class members which is situated at high
galactic latitute ($l=61.^\circ93, b=-27.^\circ 46$) at a distance of 3200\,pc,
thanks to which high-precision spectroscopic data is available and carefully
analysed (Little et al.\ 1994).

In this paper, we report the findings of our observational study of HD\,203664
in an attempt to contribute to a better understanding of $\beta\,$Cephei stellar
structure models in a diversity of such type of stars.

\section{Data description and stellar parameters}

We included HD\,203664 in the long-term photometric monitoring programme of
pulsating stars performed with the 1.2m Mercator telescope at Roque de los
Muchachos in La Palma, Canary Islands. In this framework we obtained 328 Geneva
7-colour high-precision photometric measurements between HJD\,2452085.6 and
HJD\,2452582.3. The time span of these new data is { 496.8} days.  
The integration
times were typically 4 minutes, resulting in a precision of about 7\,mmag per
measurement in U and 6\,mmag in V.  Aerts (2000) had obtained already 49
datapoints for the star between HJD\,2450391 and HJD\,2450790 with the same
instrument but attached to the 0.7m Swiss telescope at La Silla.  The Southern
and Northern Geneva standard star systems are carefully calibrated so that
measurements in both hemispheres should be compatible, even over a long
baseline.  All reduced data are provided in Table\,1.\footnote{available only in
electronic form at the CDS via anonymous ftp to cdsarc.u-strasbg.fr
(130.79.128.5)}

The basic stellar parameters of HD\,203664 were derived from different
sources. Aerts (2000) used the old Geneva photometry and positioned the star in
the HR diagram with respect to the $\beta\,$Cephei star instability strip (see
her Fig.\,1).  Using standard stellar models published by Schaller et al.\
(1992), she thus derived a mass of 13.8$M_\odot$.  We have recomputed the
estimates using the same method as in Aerts (2000) from the average value of the
6 Geneva colours for all new data and find refined values of $\log\,T_{\rm eff}=
4.47\pm 0.01$, $\log g=3.9\pm0.3$.

It is well-known, however, that fundamental parameter estimates for B stars from
high-resolution spectroscopy often result in a lower effective temperature and
gravity (see, e.g.\ De Ridder et al.\ 2004 for a discussion about this for the
$\beta\,$Cephei star $\nu\,$Eridani). Moreover, in the case of HD\,203664 we
cannot rely on an accurate value of the parallax, as shown by the large
discrepancy between the result of 0.32\,mas by Little et al.\ (1994) and of
2.23$\pm 1.08$\,mas from HIPPARCOS (Perryman et al.\ 1997).  Little et al.\
(1994) derived the stellar parameters from high-resolution spectra and used
these to estimate the distance.  They find $\log\,T_{\rm eff}= 4.447$, $\log
g=3.7$ and a mass of 14\,$M_\odot$ (based on evolutionary models by Maeder \&
Meynet 1988), as well as a normal (i.e.\ solar) abundance for B stars in our
vicinity. Unfortunately, these authors did not provide error estimates. Finally,
the few $\beta\,$Cephei stars with accurate seismic modelling have always ended
up outside their observationally determined error box in effective temperature
and gravity, to the cooler and less massive part (Thoul et al.\ 2003 for
16\,Lac, Aerts et al.\ 2003 for HD\,129929 and Pamyatnykh et al.\ 2004 and
Ausseloos et al.\ 2004 for $\nu\,$Eridani).

The final estimate of the parameters we conservatively adopt for HD\,203664,
based on all these arguments, is $\log\,T_{\rm eff}= 4.45\pm 0.02$ and $\log
g=3.8\pm 0.2$, while we do not use any constraint at all on its luminosity.

\section{Frequency analysis}
\begin{figure*}
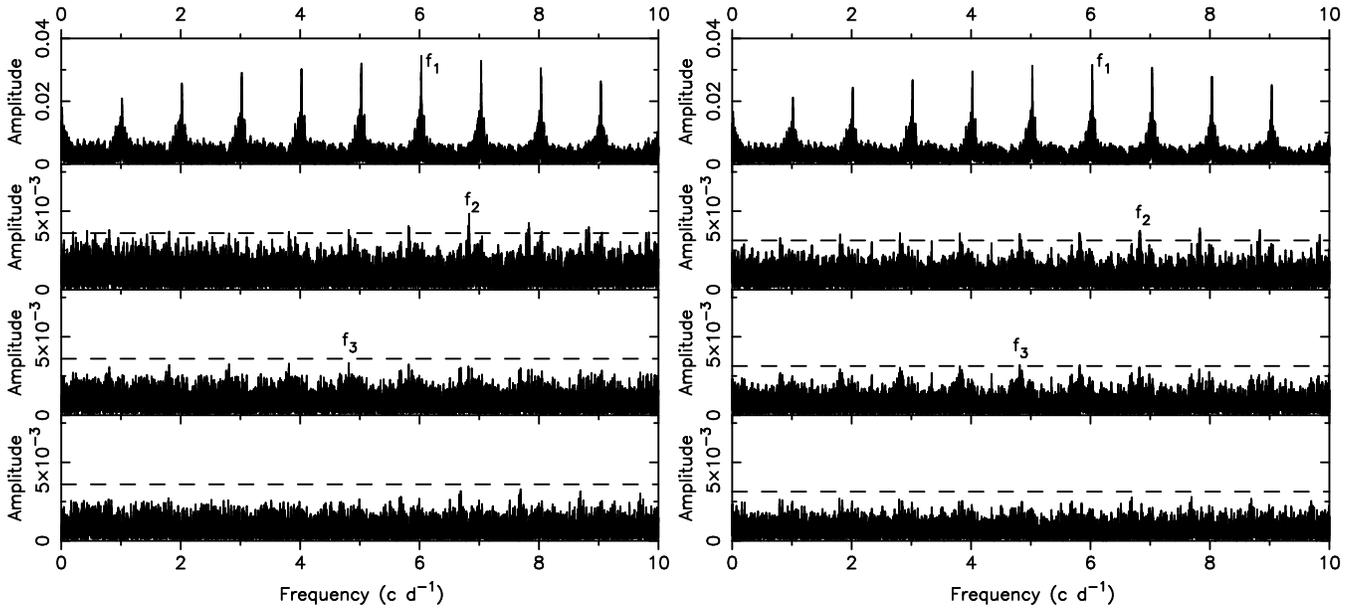

\centering
\rotatebox{-90}{\resizebox{8.cm}{!}{\includegraphics{4142.fig1a}}}
\rotatebox{-90}{\resizebox{8.cm}{!}{\includegraphics{4142.fig1b}}}
\caption{The periodograms for the Geneva U (left) and B (right) data of
HD\,203664 after subsequent stages of prewhitening with the frequencies provided
in the text. The amplitudes are expressed in magnitudes. The dashed horizontal
line indicates the 4$\sigma$ level determined in the way explained in the
text. }
\label{freqs}
\end{figure*}

We searched for frequencies in the Geneva U, B, V filters with the method
outlined in Scargle (1982). The results were similar in the B and V filter, so
we only provide the detailed analysis for the U and B filter.  We accepted
frequencies as long as their amplitude is more than 4 times the noise level,
which corresponds to a 99.9\% confidence level of having found an intrinsic
variation rather than a frequency due to noise (Breger et al.\ 1993, Kuschnig et
al.\ 1997). This criterion is common practise among asteroseismologists these
days (e.g.\ Handler et al.\ 2003, 2004, 2005). The noise level was computed by
averaging the periodogram peaks in the range $[0,10]$\,c\,d$^{-1}$ after final
prewhitening.

The accuracy of the frequencies was calculated as $\sigma_f\sim\sigma/\sqrt{N} A
T$ (Horne \& Baliunas 1986, Montgomery \& O'Donoghue 1999) where the
proportionality constant is of order 1 depending on the author (see Cuypers 1987
for a discussion), $A$ is the amplitude of the frequency $f$, $N$ is the number
of data points, $T$ is the total time base and $\sigma$ is the average error on
each individual measurement. We estimated the latter by computing the standard
deviation of the noise after final prewhitening, and found it to be 7.1\,mmag
for the Geneva U filter, 6.5\,mmag for B and 7.1\,mmag for V.

During a first step, we confirmed the dominant frequency already found by Aerts
(2000) from the HIPPARCOS photometry. This frequency appeared clearly in all
three considered filters: $f_1=6.02885(2)\,$c\,d$^{-1}$ (where the uncertainty
of the last digit is given in parenthesis). There was strong aliasing due to the
single-site nature of the data (see top panels in Fig.\,\ref{freqs}) but as the
HIPPARCOS data gave us the same value without any daily alias we were sure that
we picked the correct frequency for the dominant mode.

After prewhitening, we encountered strong frequency peaks at
n\,c\,d$^{-1}$. These peaks turned out to due to the slight difference in the
average magnitude between the older 0.70m data and the 1.2m Mercator data, which
introduces daily aliases of the yearly periods due to the observing
seasons. This small difference in average magnitude probably results from the
fact that the standard stars used for the Mercator measurements are fainter than
those used for the 0.70m Swiss telescope at La Silla.  In order not to be
disturbed by these daily aliases, we ignored the 49 older measurements in our
subsequent frequency analyses.

\begin{figure*}
\centering
\rotatebox{-90}{\resizebox{12.5cm}{!}{\includegraphics{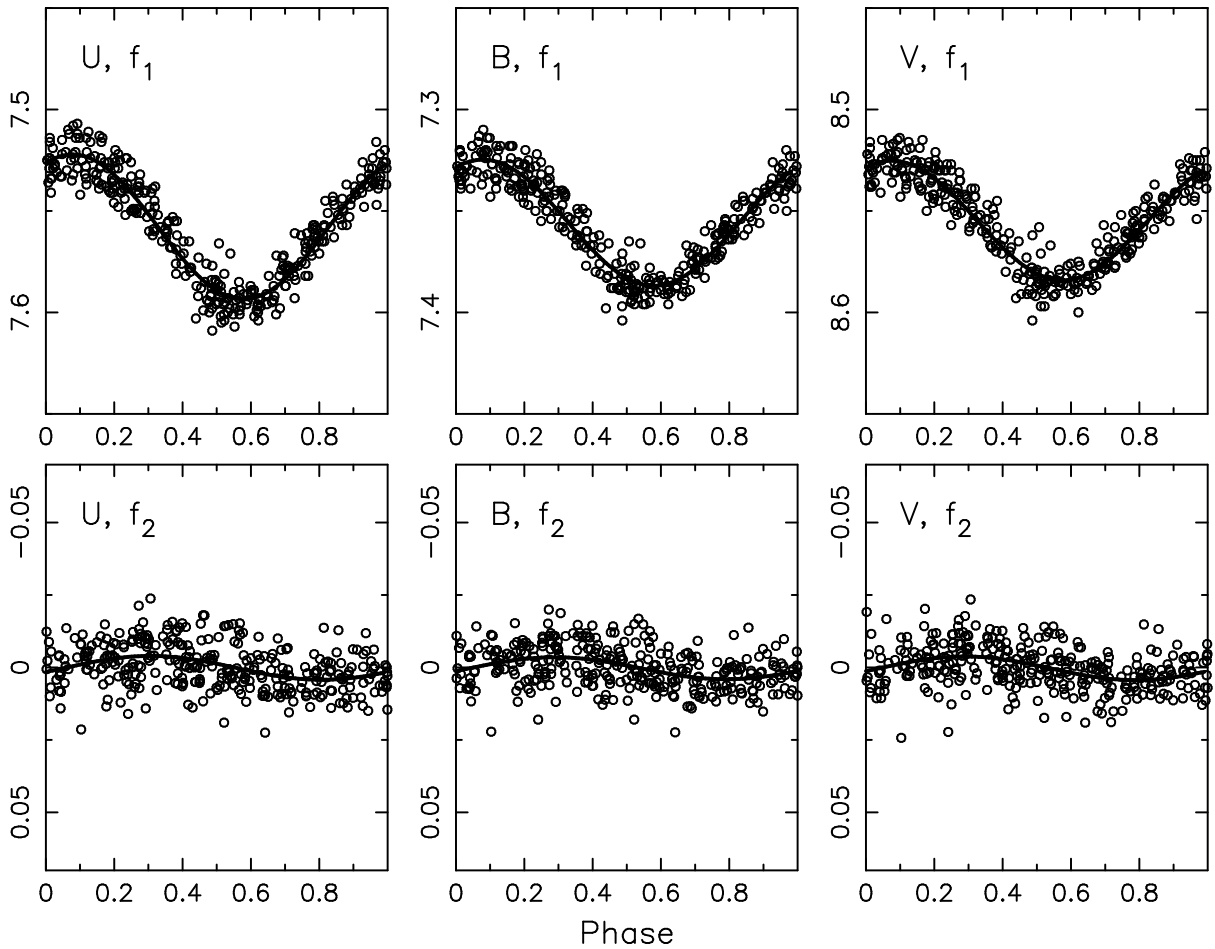}}}
\caption{Phase diagrams of the U (left), B (central), V (right) lightcurves
of HD\,203664 for $f_1$ (upper) and for $f_2$ after prewhitening with $f_1$
(lower). The dots are the data and the full line is a sinusoidal fit with fixed
indicated frequency. Note the different scale of the y-axes of the upper versus
lower plots.} 
\label{lightcurves}
\end{figure*}

Prewhitening with $f_1$ then led to a clear second frequency
$f_2=6.82902(13)\,$c\,d$^{-1}$ and its aliases in the U filter.  This frequency
and its aliases fulfilled safely the amplitude requirement
(Fig.\,\ref{freqs}). In the B filter, we found the yearly alias of $f_2$:
$f_2'=7.83176\,$c\,d$^{-1}$.  The HIPPARCOS photometry was of no use to help
discriminate among the aliases as of this stage, because there are no
significant frequencies in that dataset after prewhitening with $f_1$. We
proceeded by considering $f_2$ as well as $f_2'$ in a biperiodic fit together
with $f_1$, but this did not help to discriminate between the two options.
Therefore, we concluded that either $f_2$ or $f_2'$ is the true second
frequency. Phase plots for $f_1$ and $f_2$ for the three Geneva UBV filters are
provided in Fig.\,\ref{lightcurves}. The curves in the lower panel are
indistinguishable in quality from those for $f_2'$.

After subsequent prewhitening with either $f_2$ or $f_2'$ we continued the
search for new frequencies.  In this way, we found
$f_3=4.81543(14)\,$c\,d$^{-1}$ or one of its aliases. This frequency fulfilled
the amplitude requirement in B, but not in V and only barely in U. It must
therefore be regarded as a candidate frequency only without further
observational confirmation. Moreover, we could not distinguish the frequency
from its aliases from multiperiodic fits to the lightcurves.  Unlike for the
second frequency, numerous aliases of $f_3$ gave equally good fits.

The best overall fits to the U, B, V data was achieved by taking the values for
$f_1,\ldots,f_3$ as listed above, but we cannot exclude to have missed a true
frequency and have taken an alias for any of $f_2, f_3$. We will not use the
latter frequency to make seismic inferences further on and in the case of the
second frequency we consider each time both $f_2$ and $f_2'$.

Any other frequencies found after prewhitening do no longer reach four times the
noise level (see bottom panels of Fig.\,\ref{freqs}) so we stop the frequency
analysis at this point.

A firm conclusion is that HD\,203664 is a multiperiodic $\beta\,$Cephei star
with one dominant mode having an amplitude about ten times larger than the ones
of its other modes. The three frequencies $f_1, f_2, f_3$ reach an amplitude of
respectively 38.8, 4.5, 4.0 times the noise level in U, of 38.4, 4.7, 4.2
times the noise level in B and of 31.8, 4.5, 3.2 times the noise level in
V.

\section{Amplitude ratios and mode identification}
\renewcommand{\arraystretch}{1.2}
\begin{table*}
\caption{Results of harmonic fits to the Geneva lightcurves of HD\,203664. A
stands for the amplitude, expressed in millimag, and $\phi$ for the phase,
expressed in 2$\pi$ radians.  The adopted reference epoch for $\phi=0.0$
corresponds to HJD 2450000.0.}
\begin{center}
\begin{tabular}{ccccccccc}
\hline
&& U & B$_1$ & B & B$_2$ & V$_1$ & V & G \\
\hline
$f_1$ & A & 36.3$\pm$0.7 & 32.5$\pm$0.8 & 31.7$\pm$0.7 & 31.4$\pm$0.7 &
30.0$\pm$0.8 & 30.0$\pm$0.7 & 29.1$\pm$0.8 \\
 & $\phi$ & 0.077$\pm$0.001 & 0.078$\pm$0.001 & 0.078$\pm$0.001 & 
0.077$\pm$0.001 & 0.079$\pm$0.001 & 0.075$\pm$0.001 & 0.077$\pm$0.002 \\ 
%%%%%%%%%%%%%%%%%%%%%%%%%%%%%%%%
$f_2$ & A & 4.2$\pm$0.7 & 3.9$\pm$0.6 & 3.9$\pm$0.6 & 4.0$\pm$0.7 &
3.7$\pm$0.7 & 4.2$\pm$0.6 & 3.6$\pm$0.6 \\
 & $\phi$ & -0.210$\pm$0.026 & -0.226$\pm$0.027 & -0.226$\pm$0.024 & 
-0.199$\pm$0.028 & -0.209$\pm$0.031 & -0.225$\pm$0.023 & -0.240$\pm$0.029 \\ 
%%%%%%%%%%%%%%%%%%%%%%%%%%%%%%%%%
$f_3$ & A & 3.7$\pm$0.8 & 3.6$\pm$0.8 & 3.5$\pm$0.7 & 3.5$\pm$0.7 &
2.7$\pm$0.7 & 3.0$\pm$0.7 & 1.8$\pm$0.7 \\
 & $\phi$ & 0.172$\pm$0.010 & 0.165$\pm$0.009 & 0.166$\pm$0.009 & 
0.189$\pm$0.014 & 0.205$\pm$0.024 & 0.189$\pm$0.016 & 0.238$\pm$0.051 \\ 
%%%%%%%%%%%%%%%%%%%%%%%%%%%
\hline \multicolumn{2}{c}{variance}& 
93.2\% & 91.2\% & 92.5\% & 91.3\% & 89.3\% & 91.1\% & 87.7\% 
\\[-0.1cm] \multicolumn{2}{c}{reduction}&&&&&&&\\
\end{tabular}
\end{center}
\label{ampl}
\end{table*}
\begin{figure*}
\centering
\rotatebox{-90}{\resizebox{6cm}{!}{\includegraphics{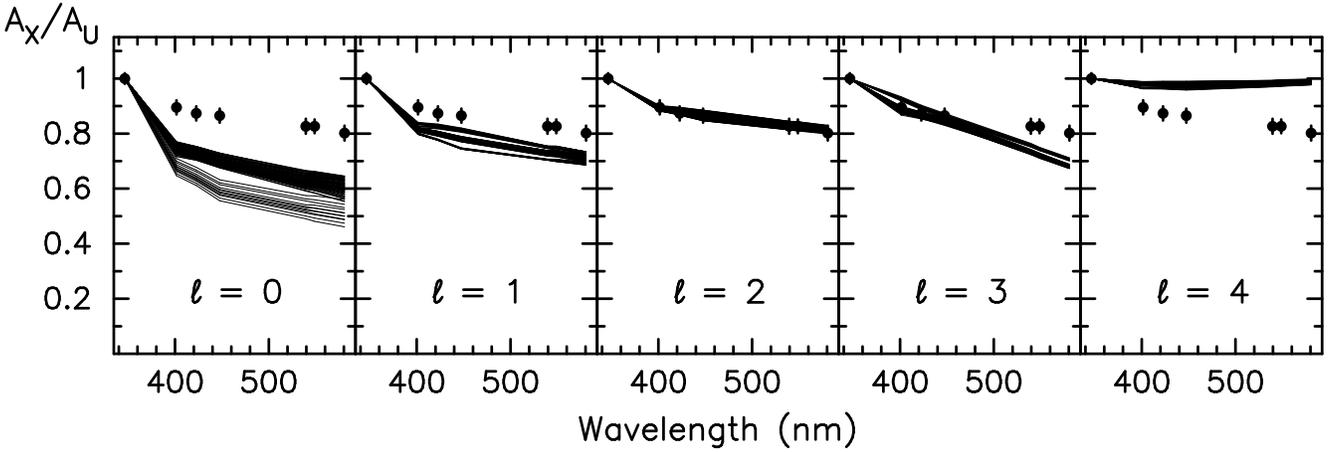}}}
\caption{Observed amplitude ratios $A_X/A_U$ (filled circles) and their
uncertainties for the dominant mode with frequency $f_1$ of HD\,203664, where
$A_X$ stands for any of the amplitudes in the seven Geneva filters U, B$_1$, B,
B$_2$, V$_1$, V, G. The lines represent theoretical predictions in the
non-adiabatic treatment of the oscillations for stellar models within the mass
range $[12,16]\,M_\odot$ during the main-sequence phase and fulfilling the error
box in $T_{\rm eff}$ and $\log g$ derived in Sect.\,2.}
\label{ratiosf1}
\end{figure*}
\begin{figure*}
\centering
\rotatebox{-90}{\resizebox{6cm}{!}{\includegraphics{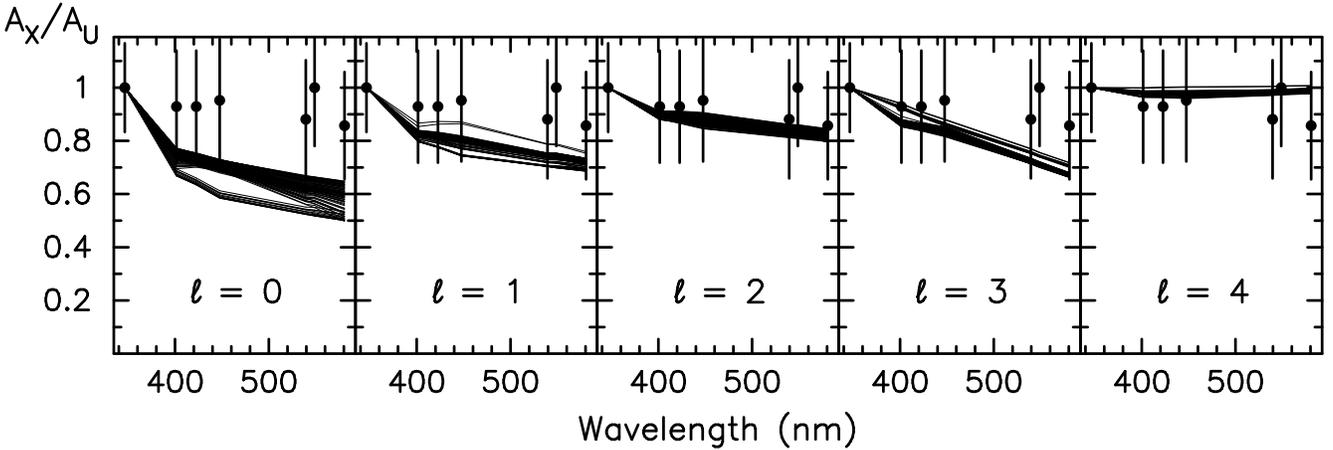}}}
\caption{Same as Fig.\,\protect\ref{ratiosf1}, but for the mode with frequency
  $f_2$.}
\label{ratiosf2}
\end{figure*}
\begin{figure*}
\centering
\rotatebox{-90}{\resizebox{12.5cm}{!}{\includegraphics{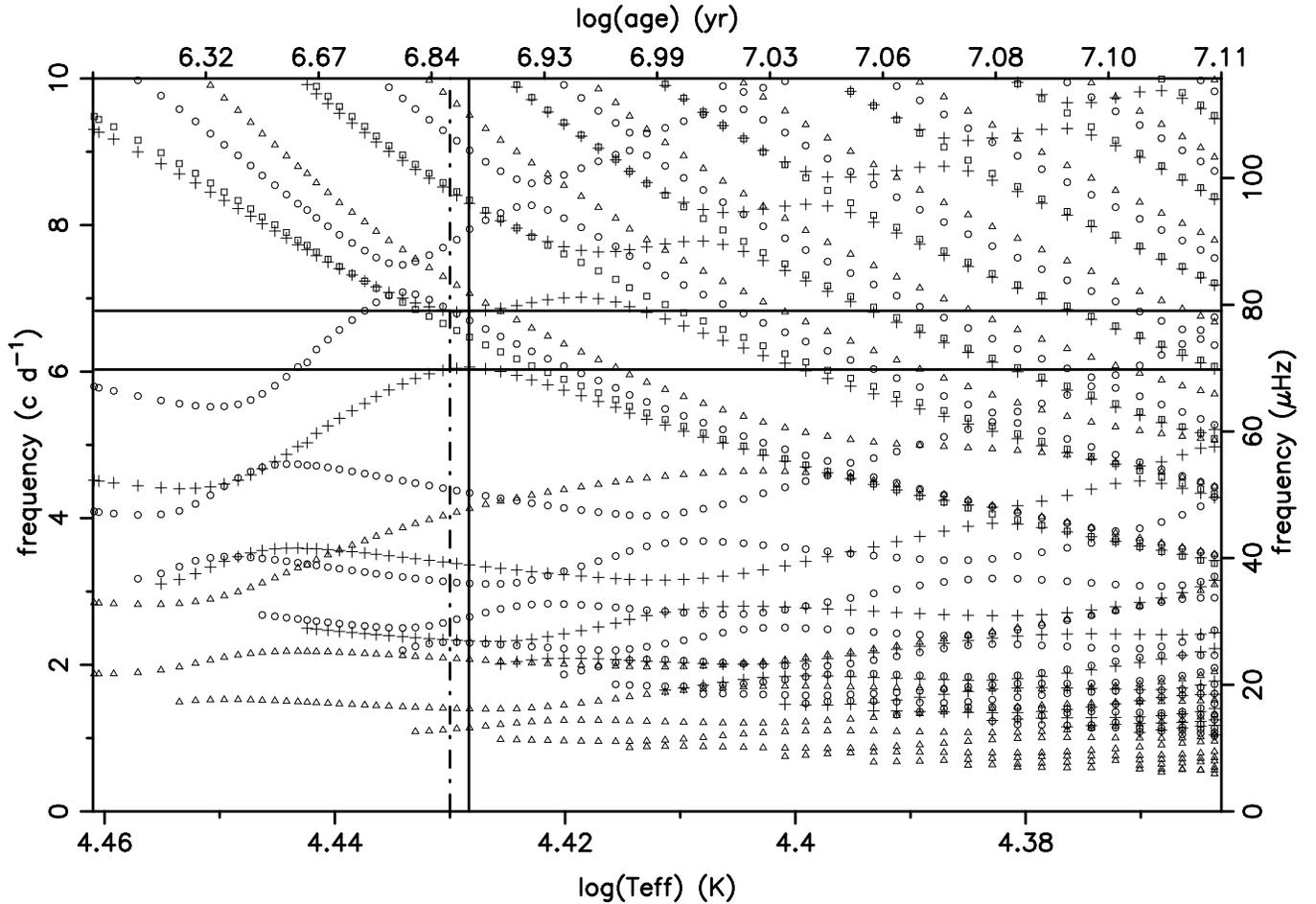}}}
\caption{{
Theoretical frequency spectra for stellar models with $M=13\,M_\odot$,
$X=0.70$, $Z=0.02$ without core overshooting from the ZAMS until the TAMS. The
star's position is to the left of the vertical dashed-dot line according to the
$T_{\rm eff}$-range derived in Sect.\,2.  The symbol convention for the modes is
as follows: squares for $\ell=0$, triangles for $\ell=1$, plus sign
for $\ell=2$, circles for $\ell=3$. Higher $\ell-$values
are not shown for clarity. The observed frequency values $f_1$ ($\ell=2$) and
$f_2$ (likely $\ell=2$) are indicated by the horizontal lines. The vertical line
indicates the model for which an exact match to $f_1$ and $f_2$ is obtained for
two zonal $\ell=2$ modes.}}
\label{scan}
\end{figure*}

Table\,\ref{ampl} lists the results for the amplitudes and phases of
least-squares fits with $f_1, f_2, f_3$ fixed at their values mentioned
above. The overall variance reduction is also listed. Using $f_2'$ rather than
$f_2$, or any alias of $f_3$ rather than these frequencies, led to amplitude and
phase values within the error bars listed in Table\,\ref{ampl}.  As a
comparison, we mention that a monoperiodic fit with $f_1$ has a variance
reduction which is typically 5 to 8\% lower depending on the filter. The
amplitudes were each time largest in the U filter, as is expected for the
oscillation modes in the $\beta\,$Cephei stars.

We observed that the phases are equal to within their accuracy for the three
modes in the seven filters. Hence, we made use only of amplitude ratios as a
mode identification diagnostic. This is according to the common procedure for
the $\beta\,$Cephei stars.

In order to attempt mode identification, we proceeded as follows. We computed
stellar models using the Code Li\'egeois d'\'Evolution Stellaire (CLES) kindly
provided by R.\ Scuflaire. For the details on the input physics, used opacity
tables and equation of state, we refer to Ausseloos et al.\ (2004).  We
restricted to models with $X=0.70$, $Z=0.02$, in agreement with the abundances
derived by Little et al.\ (1994), and without core convective overshooting.  For
each value of the mass, we computed evolutionary tracks from the ZAMS until the
TAMS and we selected those that are within the observed range of $T_{\rm eff}$
and $\log g$ derived in Sect.\,2. In doing so we considered a range in mass from
12 to 16\,$M_\odot$ in steps of 0.5M$_\odot$. Subsequently, we computed
eigenfrequencies and eigenfunctions for each of the models using the
non-adiabatic oscillation code MAD (Dupret 2001) kindly made available by M.-A.\
Dupret. For each of the models, we selected the eigenfrequency which was closest
to the measured $f_i, i=1,\ldots,3$ and considered its theoretical non-adiabatic
amplitude ratios following the formalism by Dupret et al.\ (2003). Finally, we
compared all these theoretically predicted amplitude ratios with the observed
ones for each of the detected frequencies. In this way, we performed a mode
identification which is independent of one particular stellar model, but that
considers a large range of theoretical models covering safely the current mass
estimate of HD\,203664.

The result of this procedure for the dominant mode can be found in
Fig.\,\ref{ratiosf1} for $\ell=0,\ldots,4$.  Thanks to the small error bars of
the observed ratios, we readily identified this mode as a dipole mode. Indeed,
the $\ell=2$ solution was the only one compatible with the high-quality
data.  The strongest mode of HD\,203664 is therefore clearly non-radial. A
similar situation occurs in several other $\beta\,$Cephei stars (see, e.g.,
Heynderickx et al.\ 1994).

The other two modes had too high errors on their amplitude ratios to be able to
discriminate among the $\ell$-values. We illustrate this for the mode with
frequency $f_2$ in Fig.\,\ref{ratiosf2}. While the agreement between the
theoretical predictions and the observations is best for $\ell=2$, we cannot
firmly exclude the other $\ell$-values of the non-radial modes. We do find that
the mode with frequency $f_2$ is unlikely to be radial. The same result is
obtained if we take $f_2'$ rather than $f_2$.

The ratios for the modes with $f_3$ are even more uncertain and do not
provide any constraint at all so the corresponding mode identification plot is
omitted here.

In an attempt to confront these seismic observational constraints with standard
models, we used the ones mentioned above, i.e.\ with masses ranging from 12 to
16$\,M_\odot$, $X=0.70, Z=0.02$ and without core overshooting ($\alpha_{\rm
ov}=0.0$).  We computed all their oscillation frequencies for zonal modes of
$\ell=0,\ldots,3$ { without taking into account the effects of rotation.  We
point out that this is a simplification of the true situation, because
second-order rotational effects imply frequency shifts, even for zonal
modes. Moreover, mode coupling occurs due to rotation for modes whose degree
differs by 2 if their frequencies are close together (e.g.\ Soufi et al.\ 1998,
Daszy\~nska-Daszkiewicz et al.\ 2003). However, we postpone a sophisticated
interpretation of HD\,203664's observed frequencies for the time being and
restrict to a confrontation with standard models here.}  

{ Typically three or four radial orders of an $\ell=2$ zonal mode fitting the
frequency $f_1$ resulted from the standard models from the ZAMS to the TAMS, for
each of the considered masses between 12 and 16$\,M_\odot$.  
Forcing a simultaneous fit to an $\ell=2$ zonal mode for $f_2$ was achieved for
several of the models that already fit $f_1$. An examples of the mode spectrum
of evolutionary models with $M=13\,M_\odot$ is provided in Fig.\,\ref{scan}, in
which all the $\ell=0,\ldots,3$ mode frequencies are compared with $f_1$ and
$f_2$.
The vertical dashed-dot line indicates the lower limit of the observed
effective temperature interval, i.e.\ HD\,203664 must occur to the left or close
to this line, pointing out its young age already found by Aerts (2000).  The
model indicated by the full vertical line provides a good fit to $f_1$ and $f_2$
for zonal $\ell=2$ modes. This model has parameters $X=0.70$, $Z=0.02$,
$M=13\,M_\odot$, $\log\,T_{\rm eff}$=4.429, $\log g=3.98$, $\log
L/L_\odot$=4.24, $R=6.12\,R_\odot$, a central hydrogen abundance $X_c=0.438$, an
age of $7.5\times 10^6$yr and no core overshooting. These $T_{\rm eff}$ and
$\log g$ values are compatible with the estimates from observations provided in
Sect.\,2. Moreover, non-adiabatic computations with MAD indicate that the two
modes are excited for this model. 

Fitting $f_1$ and $f_2'$ as two zonal $\ell=2$ modes is also possible if their
radial order differs by 2, but the models for which such a fit is obtained are
somewhat more evolved which is less likely due to the observational constraint
on the effective temperature.  Moreover, the frequency fit is not as good as for
$f_2$.  Any of the models with $M\neq\,13\,M_\odot$ while keeping $X$, $Z$,
$\alpha_{\rm ov}$ fixed do not provide an equally good fit to both $f_1$ and
$f_2$ (or $f_2'$) as the model indicated by the vertical line in
Fig.\,\ref{scan}.

We thus find that a good fit for $f_1$ and $f_2$ as $\ell=2$ modes can only be
achieved close to an avoided crossing {\it if we restrict to zonal modes.}
Modes near avoided crossings have very good potential to probe the interior of a
star. However, we have no conclusive observational evidence to restrict to $m=0$
and so other evolutionary models may also lead to an equally good fit of the two
considered frequencies for non-zonal modes.

At present we have too restrictive observational constraints to perform a
precise fitting procedure (e.g.\ as outlined in Ausseloos et al.\ 2004), i.e.\
to scan in detail the parameter space of any possible stellar model as a
function of $(X$, $Z$, $\alpha_{\rm ov}$, $M)$.  Indeed, besides the fact that
we cannot exclude a different $\ell-$value for the mode with frequency $f_2$ (or
$f_2'$), we have no information on the azimuthal number of the modes nor do we
know the surface rotation frequency of the star. Assuming that $f_1$ and $f_2$
represent to a good approximation the central peaks of frequency multiplets is
too restrictive because the rotational frequency must be of order
0.5\,c\,d$^{-1}$ or larger given that vsini=200\,km\,s$^{-1}$.}

We can only conclude that the currently observed oscillation spectrum of
HD\,203664 can easily be explained by standard stellar models and excitation
computations and that we have insufficient information to further constrain the
stellar parameters from asteroseismology at present.

\section{Discussion}

We have shown the large-amplitude $\beta\,$Cephei star HD\,203664 to be a
multiperiodic non-radial oscillator from single-site long-term multicolour
Geneva photometry.  The dominant mode was unambiguously identified as a dipole
$\ell=2$ mode from its amplitude ratios. The frequencies of the star are
compatible with standard stellar models of massive stars during the main
sequence phase.

There are at present nine other $\beta\,$Cephei stars known to have a
peak-to-peak V amplitude larger than the one of HD\,203664 (Stankov \& Handler
2005). Only one of these ten stars is monoperiodic (BW\,Vul) since
multiperiodicity was recently discovered in HD\,180642 (Aerts et al., in
preparation). Three of them have a confirmed dominant radial mode ($\nu\,$Eri --
Handler et al.\ 2004, BW\,Vul -- Aerts et al.\ 1995, HD\,180642 -- unpublished),
three have an $\ell=1$ mode (IL\,Vel -- Handler et al.\ 2003, 12\,Lac -- Handler
et al.\ 2005, KP\,Per -- Saesen et al., in preparation) and two have an $\ell=2$
mode (KZ\,Mus -- Handler et al.\ 2003, HD\,203664 -- this paper). We do not find
any obvious relation between these results and the projected rotation velocity
of the stars (Stankov \& Handler 2005). HD\,203664 is by far the most rapid
rotator among them, with vsini=200\,km\,s$^{-1}$ which is more than twice as
high as for any of the other large-amplitude $\beta\,$Cephei stars. From the
mass and $\log g$ estimates given in Sect.\,2, we find a critical velocity of
480\,km\,s$^{-1}$ and hence vsini / v$_{\rm crit}$ = 42\%.

The stellar parameters and internal structure of HD\,203664 can only be
constrained further by means of elimination of the alias problems presented here
for the low-amplitude modes (i.e.\ from a multisite campaign as those performed
by Handler et al.\ 2003, 2004, 2005), by subsequent unambiguous identification
of their degree from amplitude ratios and by further identification of their
azimuthal number through high-resolution spectroscopy. Given its visual
magnitude of 8.6, its oscillation periods near 4 hours and its low-amplitude
modes, together with the data requirements for such type of spectroscopic
analysis (e.g.\ Aerts \& Eyer 2000), one needs, besides a photometric multisite
campaign, at least one week of 8-m class telescope time to achieve mode
identification and subsequent interpretation of the oscillations of this massive
hot rapid rotator among the $\beta\,$Cephei stars.

\begin{acknowledgements}
We are grateful to the staff of the Geneva observatory for the generous awarding
of telescope time at La Silla and for the reduction of the data.  We thank our
colleagues from the Institute of Astronomy of Leuven University who contributed
to the gathering of the photometric data at La Silla and at La Palma. Moreover,
we are much indebted to R.\ Scuflaire, M.-A.\ Dupret and M.\ Ausseloos for the
provision of their computer source codes CLES, MAD, and SCAN respectively.
\end{acknowledgements}

{}


\begin{thebibliography}{}

\bibitem[]{} Aerts, C.\ 2000, A\&A 361, 245

\bibitem[]{} Aerts, C., Eyer, L.\ 2000, ASPC 210, 113 

\bibitem[]{} Aerts, C., Mathias, P., Van Hoolst, T., et al.\ 1995, A\&A 301, 781

\bibitem[]{} Aerts, C., Thoul, A., Daszy\'nska, J., et al.\ 2003, Sci 300, 1926

\bibitem[]{} Ausseloos, M., Scuflaire, R., Thoul, A., Aerts, C.\ 2004, MNRAS
355, 352

\bibitem[]{} Ausseloos, M.\ 2005, PhD Thesis, Katholieke Universiteit Leuven,
  Belgium 

\bibitem[]{} Breger, M., Stich, J., Garrido, R., et al.\ 1993, A\&A 271, 482

{
\bibitem[]{} Cox, A.N., Morgan, S.M., Rogers, F.J., Iglesias, C.A.\ 1992, ApJ
  393, 272}

\bibitem[]{} Cuypers, J.\ 1987, The Period Analysis of Variable Stars,
Academiae Analecta, Royal Academy of Sciences, Volume 49, Number 3, Belgium

{	
\bibitem[]{} Daszy\~nska-Daszkiewicz, J., Dziembowski, W.A., Pamyatnykh, A.A.\
  2003, ApSS 284, 133}

\bibitem[]{} De Ridder, J., Telting, J., Balona, L.A., et al.\ 2004, MNRAS 351,
  324 

\bibitem[]{} Dupret, M.-A., 2001, A\&A 366, 166

\bibitem[]{} Dupret, M-A., De Ridder, J., De Cat, P., et al.\ 2003, A\&A 398,
677

{
\bibitem[]{} Dziembowski, W.A., Pamyatnykh, A.A.\ 1993, MNRAS,
262, 204}

\bibitem[]{} Handler, G., Jerzykiewicz, M., Rodr\'{\i}guez, E., et al.\ 2005,
  MNRAS, submitted

\bibitem[]{} Handler, G., Shobbrook, R.R., Jerzykiewicz, M., et al.\ 2004, MNRAS
347, 454

\bibitem[]{} Handler, G., Shobbrook, R.R., Vuthela, F.F., et al.\ 2003, MNRAS
  341, 1005 

\bibitem[]{} Heynderickx, D., Waelkens, C., Smeyers, P.\ 1994, A\&AS, 105, 447

\bibitem[]{} Horne, J.H., Baliunas, S.L.\ 1986, ApJ 302, 757

{ 
\bibitem[]{} Kiriakidis, M., El Eid, M.F., Glatzel, W.\ 1992, MNRAS 255, L1}

\bibitem[]{} Little, J.E., Dufton, P.L., Keenan, F.P., Conlon, E.S.\ 1994, ApJ
  427, 267

\bibitem[]{} Maeder, A., Meynet, G., 1988, A\&AS, 76, 411

\bibitem[]{} Montgomery, M.H., O'Donoghue, D.\ 1999, Delta Scuti Star
  Newsletter, Volume 13

{ 
\bibitem[]{} Moskalik, P., Dziembowski, W.A.\ 2002, A\&A 256, L5}

\bibitem[]{} Pamyatnykh, A.A.\ 1999, Acta Astr., 49, 119 

\bibitem[]{}
Pamyatnykh, A.A., Handler, G., Dziembowski, W.A.\ 2004, MNRAS, 350, 1022

\bibitem[]{} Scargle, J.D.\ 1982, ApJ 263, 835

\bibitem[]{} Schaller, G., Schaerer, D., Meynet, G., Maeder, A.\ 1992, A\&AS,
  96, 269

{
\bibitem[]{} Soufi, F., Goupil, M.J., Dziembowski, W.A.\ 1998, A\&A 334, 911}

\bibitem[]{} Stankov, A., Handler, G.\ 2005, ApJS 158, 193

\bibitem[]{} Thoul, A., Aerts, C., Dupret, M.-A., et al.\ 2003, A\&A 406, 287

\bibitem[]{} Waelkens, C., Aerts, C., Kestens, E., Grenon, M., Eyer, L.\ 1998,
    A\&A 330, 215

\end{thebibliography}
\end{document}